\documentclass[twoside]{article}

\usepackage[OT1]{fontenc} 
\usepackage[latin1]{inputenc}
\usepackage{graphicx}
\usepackage{actes}
\usepackage[francais]{babel}
\usepackage{subfigure}

\usepackage{relsize}

\usepackage{amsmath}
\usepackage{amssymb}
\usepackage{stmaryrd}

\usepackage{xspace}
\usepackage{url}
\usepackage{boxedminipage}
\newcommand{\inter}{\cap}
\newcommand{\union}{\cup}

\newcommand{\bottom}{\bot}

\newcommand{\regle}{\cfrac}

\newcommand{\lcroch}{\llbracket}
\newcommand{\rcroch}{\rrbracket}

\newcommand{\implique}{\ensuremath{\Rightarrow}\xspace}
\newcommand{\fleche}{\rightarrow}

\newcommand{\stype}{\subseteq}

\newcommand{\rem}[1]{}
\newcommand{\mcal}{\mathcal}

\newcommand{\infere}{\vdash}

\newcommand{\col}[2]{[#2]#1}

\usepackage{theorem}
\theorembodyfont{\normalfont}
\newtheorem{lemme}{Lemme}
\newcommand{\Vois}{\ensuremath{\mathit{Vois}}}
\newcommand{\Sol}{\ensuremath{\mathit{Sol}}}
\newcommand{\Comp}{\ensuremath{\mathit{Comp}}}
\newcommand{\est}{\ensuremath{\mathit{est}}}
\newcommand{\nord}{\ensuremath{\mathit{nord}}}
\newcommand{\nordnb}{\ensuremath{\mathit{nord\_nb}}}
\newcommand{\seq}{\ensuremath{\mathit{seq}}}
\newcommand{\set}{\ensuremath{\mathit{set}}}
\newcommand{\bag}{\ensuremath{\mathit{bag}}}
\newcommand{\grid}{\ensuremath{\mathit{grid}}}
\renewcommand{\int}{\ensuremath{\mathit{int}}}
\newcommand{\bool}{\ensuremath{\mathit{bool}}}
\newcommand{\float}{\ensuremath{\mathit{float}}}
\newcommand{\empt}{\ensuremath{\mathit{empty}}}
\newcommand{\wrong}{\ensuremath{\mathit{wrong}}}
\newcommand{\shaperr}{\ensuremath{\mathit{shape\_err}}}
\newcommand{\self}{\ensuremath{\mathit{self}}}
\newcommand{\Eval}{\ensuremath{\mathit{Eval}}}

\newcommand{\fort}{\ensuremath{\mathcal L _=}}
\newcommand{\souple}{\ensuremath{\mathcal L _\subseteq}}

\title{Typage fort et typage souple des collections topologiques et
des transformations
}

\author{Julien Cohen}

\titlehead{Typage des collections topologiques et des transformations}%  a droite (page impaire)
\authorhead{J. Cohen}% a gauche (page paire)

\affiliation{\begin{tabular}{rr} 
\\   LaMI  UMR 8042,
\\   CNRS - Université d'\'Evry Val d'Essonne
\\   523 Place des Terrasses de l'Agora
\\	91025 \'Evry, France  
\\     {\tt jcohen@lami.univ-evry.fr} 
\end{tabular}}

\begin{document}
\setcounter{page}{1}
\maketitle
%

% INTRO %%%%%%%%%%%%%%%%%%%%%%%%%%%%%%%%%%%%%%%%%%%%%%%%%%%%%%%%%%%%%%%%%%%%

\begin{abstract}
Les collections topologiques permettent de considérer uniformément de
nombreuses structures de données dans un langage de programmation et
sont manipulées par des fonctions définies par filtrage appelées des
transformations.

Nous présentons dans cet article deux systèmes de types pour des
langages intégrant les collections topologiques et les
transformations. Le premier est un système à typage fort à la
Hindley/Milner qui peut être entièrement typé à la compilation. Le
second est un système à typage mixte statique/dynamique permettant
de gérer des collections hétérogènes, c'est-à-dire qui contiennent des
valeurs de types distincts. Dans les deux cas l'inférence de types
automatique est possible.

\end{abstract}

\section{Introduction}
\label{intro}

Les \emph{collections topologiques} sont une famille de structures de
données que l'on peut voir comme des fonctions d'un ensemble de
positions vers un ensemble de valeurs et sur lesquelles on dispose
d'une relation de voisinage entre les positions. De nombreuses
structures de données usuelles peuvent être vues comme des collections
topologiques~: ensembles, séquences, tableaux généralisés, graphes,
\emph{etc}.  On peut programmer par filtrage sur ces collections grâce
à des fonctions particulières appelées \emph{transformations}. Les
transformations permettent d'écrire des programmes opérant
uniformément sur diverses structures de données. De tels programmes
sont dits \emph{polytypiques}~\cite{polytypism}.

Nous montrons dans cet article que les collections topologiques et les
transformations peuvent s'intégrer dans un langage fortement typé. De
plus elles y apportent des caractéristiques importantes comme le
polytypisme et le filtrage sur des structures non-algébriques sans
perdre le polymorphisme paramétrique ou l'inférence automatique de
types.

Dans un second temps, nous proposons un système de typage plus souple
permettant de manipuler des collections hétérogènes, c'est à dire des
collections contenant des valeurs de types différents. En effet, la
programmation par transformations trouve un champ d'application
important en simulation biologique où les collections manipulées sont
souvent hétérogènes\rem{ et où un langage au typage souple est donc
plus proche des besoins des utilisateurs}. Le cadre hétérogène permet
aux transformations de gagner en expressivité par rapport au langage
fortement typé. Des exemples de tels programmes~\cite{bugrim,
meristem} ont été codés dans le langage déclaratif MGS~\cite{rta03}
qui intègre les collections et les transformations dans un contexte
dynamiquement typé.  Les langages dynamiquement typés ont fait l'objet
de nombreux travaux visant à leur donner un système de types proche
d'un typage statique avec inférence de types automatique~\cite{aw94,
fagan, Damm3, furuse}. Le langage que nous présentons qui permet la
manipulation de collections hétérogènes est un sous-ensemble du
langage MGS et nous lui associons un système de types souple en partie
basé sur les travaux de Aiken~\emph{et al.}~\cite{aw94}. Ce système
est doté d'une procédure de typage automatisée permettant de
construire un type précis pour les transformations. Ce système de
types est une étape importante dans l'élaboration d'un compilateur
efficace pour le langage MGS et peut s'adapter à d'autres langages à
base de règles de réécriture.

\hyphenation{to-po-lo-gi-que}

La section suivante donne une intuition du fonctionnement des
collections topologiques et des transformations et introduit leur
typage. La section~\ref{fort} présente notre langage fortement typé et
son système de types. La section~\ref{souple} présente le langage à
typage souple et énonce la correction de ce typage puis donne un
aperçu de l'intérêt du typage pour la compilation des transformations.
La dernière section conclut cet article en proposant des extensions
directes de nos travaux et en présentant les travaux proches des
nôtres et les perspectives ouvertes par notre travail.

% COLLECTIONS ET TRANSFORMATIONS %%%%%%%%%%%%%%%%%%%%%%%%%%%%%%%%%%%%%%%%%%%%%
\section{Collections topologiques et transformations}
\label{intuition}

Dans cette section nous introduisons les collections topologiques puis
les transformations de manière informelle. Nous donnons également des
éléments pour comprendre comment un système de types pour un langage
fonctionnel peut les intégrer.

\subsubsection*{Collections topologiques}

Une collection topologique est une structure de données sur laquelle
il existe une relation de voisinage notée~{\Vois} entre les éléments.
Lorsqu'on a~$\Vois(e_1,e_2)$ on dira que~$e_2$ est un voisin de~$e_1$.
Par exemple, une \emph{séquence} est une collection topologique telle que~:\begin{itemize}

\item
chaque élément possède au plus un voisin~;

\item
chaque élément ne peut être le voisin que d'un élément au plus~;

\item
il n'existe pas de cycle dans la relation de voisinage.

\end{itemize}

 Un ensemble ou un multi-ensemble peuvent être vus comme une
collection dont tout élément est voisin de tous les autres éléments.

La \emph{grille} est un autre exemple de collection topologique qui
est similaire à une matrice. Chaque élément contenu dans une grille
peut avoir quatre voisins se trouvant respectivement au nord, au sud,
à l'est et à l'ouest. Contrairement aux tableaux,
la grille est une structure de données partielle.

%\medskip
\begin{figure}[h]

\subfigure[séquence]{
\includegraphics[width=2.5cm]{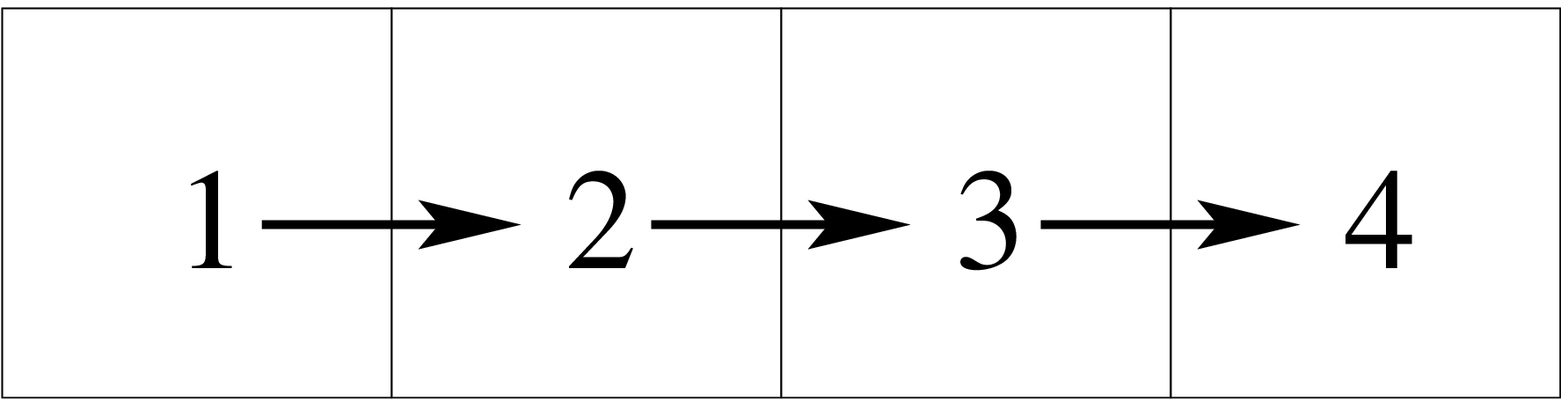}
}
\hfill
\subfigure[ensemble]{
\includegraphics[height=2.1cm]{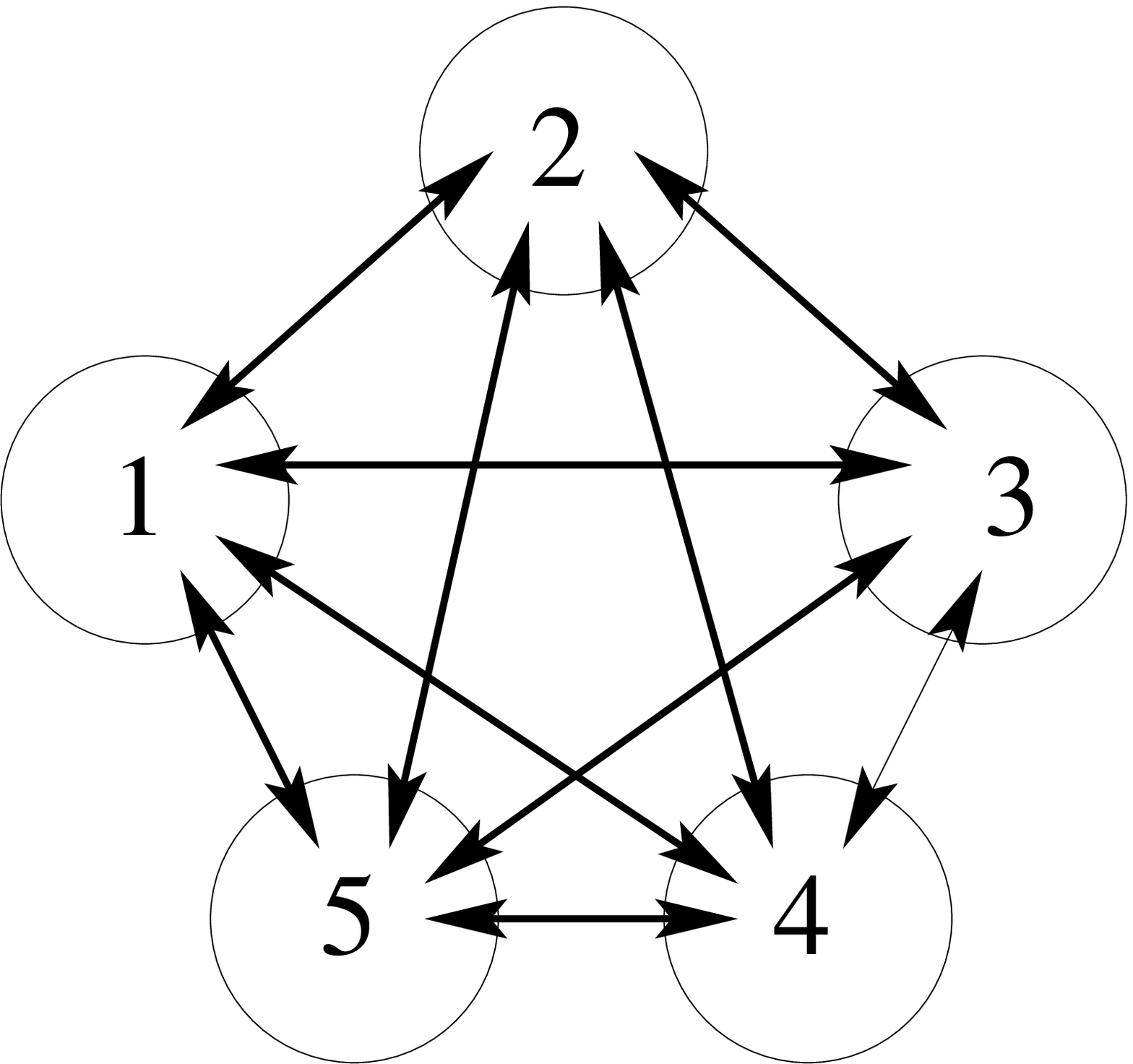}
}
\hfill
\subfigure[grille]{
\includegraphics[height=2.1cm]{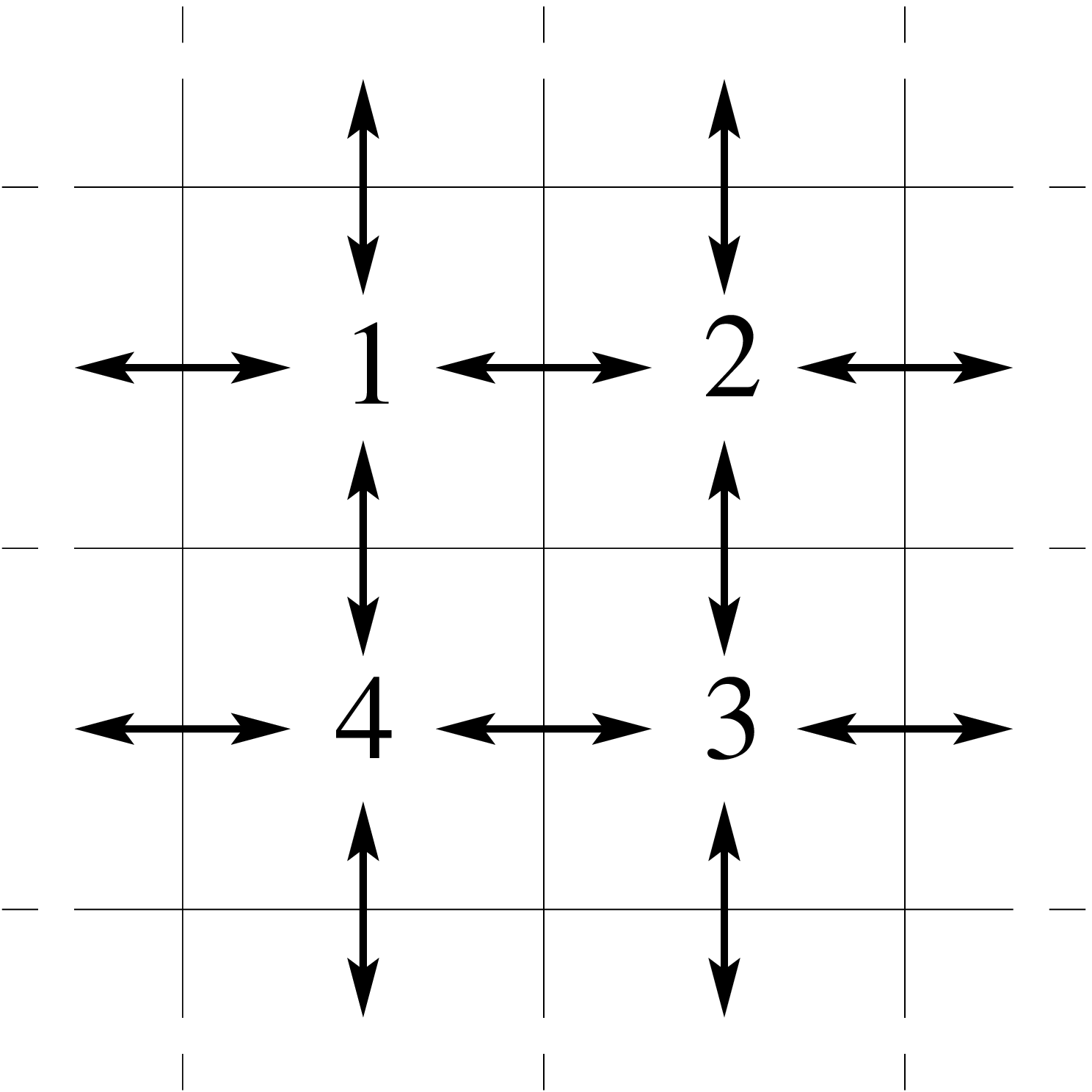}
}

\caption{exemples de collections topologiques}
\end{figure}
%\medskip

De nombreuses structures de données peuvent être considérées
uniformément comme des collections topologiques, aussi bien des
structures usuelles comme celles que nous venons de décrire que des
structures plus spécialisées comme les graphes de Delaunay.\\

Dans le langage que nous proposons les collections peuvent être
construites à partir de collections vides et d'opérateurs de
construction. Par exemple $1::\empt\_\set$ produit l'ensemble
contenant l'entier~$1$. L'opérateur $::$ peut être utilisé pour
construire n'importe quelle sorte de collection\footnote{La sémantique
de l'opérateur~$::$ dépend de la topologie de la collection à laquelle
il est appliqué. Cette forme de surcharge est de la même nature que la
surcharge du~$=$ de ML.}, ainsi on peut l'utiliser pour produire une
séquence comme dans $1::2::\empt\_\seq$.

Quatre opérateurs sont réservés à la construction de grilles :
$\nord$, $-\nord$, $\est$ et~$-\est$. Ceux-ci 
permettent de spécifier l'organisation entre les éléments lors de la
construction de la collection. Ainsi on peut construire une grille
carrée~: $1~ \est~ 2~ \nord~ 3~ \mbox{$-\est$}~ 4~ ::~ \empt\_\grid$ ou une grille
triangulaire~: $1~ \mbox{$-\nord$}~ 2~ \est~ 3~ ::~ \empt\_\grid$ par exemple.

\begin{center}
\includegraphics[width=5cm]{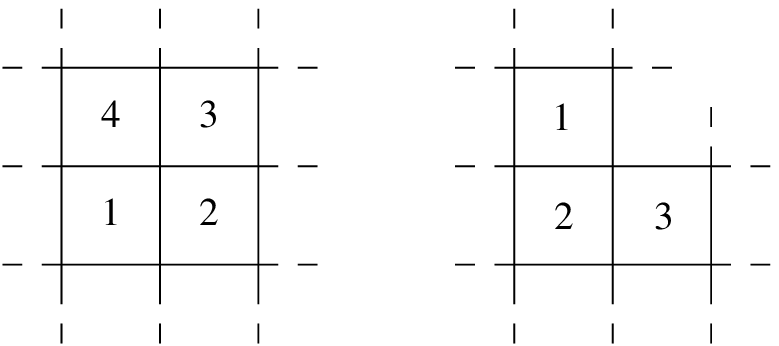}
\end{center}

Les systèmes de types que nous proposons dans cet article contiennent
des types particuliers de la forme~$\col\rho\tau$ pour les collections
où~$\tau$ est le type des éléments contenus dans la collection aussi
appelé le \emph{type contenu} de la collection et~$\rho$ est sa
\emph{topologie}. Une topologie peut être soit un symbole de
l'ensemble $\{\set,~ \bag,~ \seq,~ \grid,~\dots\}$ qui représente les
topologies possibles des collections, soit une variable de topologie
que l'on pourra dénoter par la lettre grecque~$\theta$. Notons qu'une
topologie n'est pas un type et qu'une variable de topologie ne peut
être utilisée à la place d'une variable de type et \emph{vice versa}.
Les deux grilles ci-dessus par exemple ont le type
$\col{\grid}{\int}$. Le constructeur $\nord$ et les trois autres
constructeurs spécifiques aux grilles ont le type $\alpha
\fleche \col{\grid}{\alpha} \fleche
\col{\grid}{\alpha}$ où~$\alpha$ est une variable de type.
Le constructeur générique~$::$ a quant à lui le type $\alpha\fleche
\col{\theta}{\alpha} \fleche \col{\theta}{\alpha}$ car il peut être
utilisé avec toute collection, quelle que soit sa topologie.\\

On parle de collections \emph{hétérogènes} lorsque les valeurs
contenues dans les collections peuvent être de types
différents. L'ensemble $\{1, ~true\}$ contient deux valeurs de types
respectifs $\int$ et $\bool$ et est donc un exemple de collection
hétérogène. Le langage fortement typé ne permettra pas de manipuler
des collections hétérogènes.

Pour rendre compte de l'hétérogénéité des collections dans notre
 système de types souple nous utilisons des types~\emph{unions}, déjà
utilisés par d'autres auteurs~\cite{aiken93, fol, Damm1,
frisch02semantic}. Une valeur du type union~$\tau_1 \union \tau_2$ est
soit du type $\tau_1$ soit du type $\tau_2$. Savoir qu'une valeur est
du type $\tau_1\union\tau_2$ ne permet pas de déduire qu'elle
est du type~$\tau_1$. L'entier $1$ du type $\int$ est aussi du type
$\int \union \bool$.

Nous pouvons à présent donner un type à l'ensemble~$\{1,~true\}$, ce
type est~\mbox{$\col{\set}{\int\union \bool}$}. On peut lire ce type de
la manière suivante~: { \og collection de topologie $\set$ qui contient des
valeurs de type~$\int$ et des valeurs de type~$\bool$ \fg } mais il est plus
juste de le comprendre ainsi~: { \og collection de topologie $\set$ dont
les éléments ont le type $\int
\union \bool$ \fg }. 

L'utilisation des collections hétérogènes nécessite la possibilité de
tester le type des valeurs à l'exécution. Ceci explique que le langage
les manipulant ne puisse être entièrement typé à la compilation.\\

\subsubsection*{Transformations}

Une transformation est une fonction opérant sur les collections
définie par un ensemble de règles de réécriture de la forme
$m\implique e$ appelées simplement~\emph{règles}. La partie
gauche d'un règle est appelée~\emph{motif} et la partie droite
\emph{expression de remplacement}. On note une transformation par
l'énumération de ses règles entre accolades~: $\{m_1\implique e_1
;\dots; m_n\implique e_n\}$.

L'application d'une transformation à une collection se fait en
 appliquant les règles de la transformation de la manière suivante~:
 des instances disjointes du motif de la première règle sont
 recherchées puis, lorsqu'on ne peut plus trouver de nouvelle instance
 du motif on recherche des instances du motif de la seconde règle
 parmi les éléments qui n'ont pas déjà été filtrés, et ainsi de
 suite. Lorsque ce processus de filtrage est terminé on substitue les
 parties filtrées par les parties remplaçantes correspondantes et la
 nouvelle collection ainsi créée est retournée.

Le motif $x$ filtre une valeur quelconque~; le motif $x:\int$ filtre
une valeur de type~$\int$ (uniquement dans le langage à typage
souple)~; le motif gardé $x/p(x)$ filtre une valeur $v$ telle que
$p(v)$ s'évalue à $true$~; le motif $x,y$ filtre deux valeurs voisines
quelconques~; enfin le motif $x\!:\!\int,~ y\!:\!\int ~/~x\!>\!y$
filtre deux valeurs entières voisines telles que la première est plus
grande que la seconde.  Le processus de filtrage a été décrit
formellement dans~\cite{rule02}.

Les transformations permettent d'exprimer simplement des programmes
classiques comme le tri d'une séquence par exemple. 
Pour trier une séquence on peut chercher des couples d'éléments
voisins mal ordonnés dans la séquence et les placer dans le bon
ordre. Lorsqu'il n'existe plus de couples d'éléments voisins mal
ordonnés la séquence est triée. Cette sorte de tri à bulles est
obtenue en itérant l'application de la transformation suivante jusqu'à
atteindre un point fixe~: $ \{ x,y /x>y \implique [y,x]\}$.

Dans cet exemple, l'expression $[y,x]$ en partie droite de la règle
dénote une séquence à deux éléments. En effet, lors du processus de
filtrage, une instance du motif $x,y$ est représentée par une suite de
deux valeurs se trouvant en des positions voisines dans la
collection. Cette suite de valeurs positionnées est appelée
un~\emph{chemin} dans la collection. Ici le chemin est de longueur~2.
 Les valeurs dénotées par la séquence $[y,x]$ viendront remplacer les
valeurs du chemin filtré par la partie gauche de la règle.

L'utilisation d'une séquence comme expression de remplacement convient
quelle que soit la topologie de la collection à laquelle la
transformation est appliquée. En effet, un motif filtre un chemin qui
peut être vu comme une séquence de valeurs positionnées. Une séquence
de valeurs est donc suffisante en partie droite pour spécifier le
remplacement point à point des éléments filtrés. C'est pourquoi nous
forçons les parties droites de règles à être des séquences.

Dans certaines collections comme les grilles la longueur de la
séquence remplaçante doit être égale à la longueur du chemin filtré et
la substitution se fera point à point. Dans le cas contraire, la
topologie de la collection ne serait pas préservée~: une valeur ne
peut être remplacée par plusieurs valeurs dans une grille car il
faudrait pour cela insérer de nouvelles positions et la collection ne
serait plus une grille.  Ces collections dont l'ensemble de positions
n'est pas modifiable sont dites \emph{newtoniennes}\footnote{Cette
appellation vient de la vision différente de la notion d'espace selon
Leibnitz ou Newton~\cite{rta03}.}. En revanche dans une séquence, un
ensemble ou un multi-ensemble le chemin filtré peut être remplacé par
un nombre arbitraire d'éléments car on peut toujours insérer une
position entre deux positions dans ces collections. Ces collections
dont l'ensemble de positions peut varier sont dites
\emph{leibnitziennes}.\\

%début map
La fonction $map$ qui applique une fonction à tout élément d'une
collection est un autre exemple de programme simple à écrire~:
$\lambda f.\lambda c.(let~t=\{x\implique [f\;x]\}~in~(t~c))$ ou de
manière équivalente $\lambda f.\{x\implique [f\;x]\}$ . Ceci
implémente bien un map car chaque élément $e$ de la collection sera
filtré par le motif $x$ et sera remplacé par $f(e)$. Cette fonction
peut s'appliquer à toute collection, indépendamment de sa
topologie. De telles fonctions sont dites
\emph{polytypiques}~\cite{polytypism}. Le polytypisme est l'un des
avantages à considérer les structures de données dans un cadre unificateur.
%fin map

L'identité sur les collections peut s'écrire $\{x \implique [x]\}$ et
a le type $\col\theta\alpha \fleche \col\theta\alpha$. En effet cette
transformation s'applique à toute collection topologique et ne change
ni sa topologie, ni son type contenu. De manière générale une
transformation ne change pas la topologie de la collection à laquelle
elle est appliquée. La transformation $\{x:\int \implique[x+1]\}$ où
$+$ est l'addition entière a également le type
$\col\theta\alpha\fleche \col\theta\alpha$ (dans le typage souple). En
revanche la transformation $\{x \implique [x+1]\}$ a le type
$\col\theta{\int}\fleche \col\theta{\int}$ dans les deux systèmes car une erreur de type aura
lieu si la règle est appliquée à une valeur qui n'est pas du type~$\int$.

Considérons à présent la transformation $\{x:\int ~/~(x~mod~2=0)
\implique [true]\}$ dans le langage à typage souple. Celle-ci est du
type $\col\theta\alpha \fleche \col\theta{\alpha\union \bool}$ mais ce
type manque de précision.  Il ne porte pas l'information que des
booléens apparaissent dans la collection renvoyée uniquement si la
collection en argument contient des entiers. Pour traduire cette
information nous utiliserons des types \emph{conditionnels} de la
forme~$\tau_1?\tau_2$. Le type~$\tau_1?\tau_2$ se lit { \og $\tau_1$
\emph{if} $\tau_2$ \fg} et vaut $\tau_1$ lorsque $\tau_2$ n'est pas égal
au type nul noté~$0$ et il vaudra~$0$ sinon. Ainsi on peut donner le
type $\col\theta\alpha \fleche\col\theta{\alpha\union (\bool?(\alpha
\inter \int))}$ à la transformation mentionnée ci-dessus. Ce type
signifie que si la collection en argument contient des valeurs de
type~$\int$ alors la collection renvoyée pourra contenir des valeurs de
type~$\bool$ car dans ce cas $\alpha \inter \int$ n'est pas nul et
vaut~$\int$. En revanche si le type de la collection en argument
indique qu'elle ne contient pas de valeurs du type~$\int$ alors le
type de la collection renvoyée est le même que le type de la
collection en argument. En effet dans ce cas $\alpha \inter \int$ sera
nul et donc $\bool?(\alpha\inter \int)$ sera également nul et
$\alpha\union \bool?(\alpha\inter \int) $ vaudra~$\alpha$.

Le typage fin des transformations avec des types conditionnels peut
être vu comme de l'analyse de flots. On peut noter que les types
conditionnels ont déjà été utilisés à cette fin
(voir~\cite{control_flow}).

\rem{Cet exemple montre que le système de types que nous présentons permet
d'obtenir une bonne précision et peut être vu comme une analyse de
flot de contrôle. Les types conditionnels ont d'ailleurs été utilisés
dans un but d'analyse de flot de contrôle de programmes fonctionnels
dans~\cite{control_flow}.}

% TYPAGE FORT %%%%%%%%%%%%%%%%%%%%%%%%%%%%%%%%%%%%%%%%%%%%%%%%%%%%%%%%%%%%%%%%%%%

\section{Typage fort}
\label{fort}
Le premier langage que nous présentons, noté~$\mathcal L_=$ permet de
manipuler des collections homogènes et des transformations. Il a pour
objet de montrer que les collections topologiques et les
transformations s'intègrent bien dans un langage fortement typé comme
ML. Nous présentons donc un système de types pour~$\mathcal L_=$ qui
permet l'inférence automatique des types par une extension de
l'algorithme de Damas/Milner.

\subsection{Langage $\mcal L_=$}%%%%%%%%%%%%%%%%%%%%%%%%%%%%%%%%%%%%%%%%%%%
\label{langage-fort}

Le langage $\mcal L_=$ est un $\lambda$-calcul avec constantes et
$let$ auquel on ajoute les transformations. 

\begin{center}
\begin{tabular}{lrl}
$e$ & ::= & $x~|~c ~|~ \lambda x . e  ~|~ e~e ~|~ let~x~=~e~in~e ~|~
\{ m/e \implique e; \dots ;  m/e \implique e; x \implique e\}$\\
& & \\
$ m $ 	& ::= & $x,~\dots~,~x$
\end{tabular}
\end{center}

Une transformation est composée d'une suite de règles dont la dernière
dispose d'un motif qui se réduit à une variable. Une telle règle de la
forme~$x\implique e$ est appelée \emph{attrape-tout}. Cette dernière
règle permet de garantir que toutes les valeurs d'une collection
seront filtrées par la transformation. Ainsi si les règles d'une
transformation remplacent des entiers par des flottants par exemple on
est sûr que tous les entiers seront remplacés et on peut garantir que
la transformation préserve l'homogénéité des collections.

Parmi les constantes du langage on trouve notamment des collections
vides ($\empt\_\seq$, $\empt\_\grid$, \dots) et des opérateurs comme le
constructeur générique de collections noté $::$ ou le constructeur
spécialisé~$\nord$.

Nous ferons un grand usage de sucre syntaxique, notamment~:
\begin{itemize}

\item
les opérateurs binaires seront écrits en position infixe,

\item
on pourra écrire une séquence en énumérant ses éléments entre crochets au lieu d'utiliser le constructeur standard comme le montre l'exemple suivant~: $[1,2,3]$ pour $1::2::3::\empt\_\seq$,

\item
on pourra omettre la garde d'un motif lorsque celle-ci est la constante $true$.
\end{itemize}

\rem{
Le domaine dans lequel le langage est évalué est noté~$D$ et il
contient les collections topologiques en plus des valeurs
usuelles. Nous ne détaillons pas la construction de $D$ ici, on pourra
se référer à~\cite{topological2002} pour le modèle des collections
topologiques. Comme usuellement, $wrong$ est une valeur particulière
de~$D$ renvoyée en cas d'erreur de type, par exemple lorsqu'une valeur
n'étant pas une fonction est appliquée. Une autre valeur particulière
$shape\_err$ existe dans $D$ pour rendre compte des erreurs pouvant
survenir lorsque l'application d'une règle changerait la topologie
d'une collection newtonienne. 
La fonction $eval$ qui donne la sémantique d'une expression n'est pas
détaillée non plus. Toutefois on peut trouver dans~\cite{tech89} une
définition de~$eval$ sur les transformations qui est valide pour les
deux langages que nous présentons ici.  On peut noter que les
transformations sont des fonctions et n'ont donc pas besoin d'une
construction qui leur soit propre dans $D$.}

Nous ne donnons pas ici la sémantique formelle de \fort. Le lecteur
pourra se référer à~\cite{topological2002} ou à~\cite{tech61} pour le
modèle des collections topologiques et à~\cite{tech89} pour la
sémantique des transformations.
Deux valeurs particulières dénotent des erreurs : $\wrong$ lorsqu'une
erreur de type survient et $\shaperr$ lorsque l'application d'une règle
viole la topologie d'une collection newtonienne.

\subsection{Types}%%%%%%%%%%%%%%%%%%%%%%%%%%%%%%%%%%%%%%%%%%%%%%%%%%%
\label{types-fort}

Nous associons à $\mathcal L_=$ un système de types à la
Hindley/Milner augmenté des types collections.  L'ensemble des
topologies possibles pour une collection topologique est
noté~$R$ et contient au moins $\set$, $\seq$, $\bag$ et
$\grid$. On note~$B$ l'ensemble des types de base $\{\int,~\bool,~\float,~string\}$.
$$\tau ::= \alpha ~|~ B ~|~ \tau \fleche \tau ~|~ \col\rho\tau
\hspace{2cm}
\rho ::=  \theta ~|~ R $$
La lettre $\theta$ sera utilisée pour désigner une variable de
topologie.

Un schéma de types~$\sigma$ est un type quantifié sur des variables de type et
des variables de topologie de la forme~$\forall \alpha_1,
\dots,\alpha_n,\theta_1, \dots, \theta_m . \tau$. On dit qu'un
type~$\tau$ est une instance d'un schéma de types~$\sigma=\forall \alpha_1,
\dots,\alpha_n,\theta_1, \dots, \theta_m . \tau'$ s'il existe
une instanciation~$s$ des variables quantifiées de~$\sigma$ telle
que~$s(\tau') = \tau$ et on le note~$\sigma \leq \tau$.

La fonction $TC$  donne le schéma de type associé aux
constantes du langage. Par exemple $TC(::)$ vaut
$\forall\alpha,\theta. \alpha \fleche \col\theta\alpha \fleche
\col\theta\alpha$.

Un \emph{contexte} de typage~$\Gamma$  est une fonction d'un ensemble de
variables du langage vers l'ensemble des schémas de types.

\subsubsection*{Règles de typage}%%%%%%%%%%%%%%%%%%%%%%%%%%%%%%%%%%%%%%%%%%%%%%

Les règles d'inférence sont celles d'Hindley/Milner augmentées d'une
règle pour les transformations.

$$\regle{
\Gamma(x) \leq \tau 
}{
\Gamma \infere x : \tau}~(var-inst)
\quad\quad
\regle{
TC(c) \leq \tau
}{
\Gamma \infere c:\tau}~(const-inst)
$$

$$\regle{
\Gamma\union\{x:\tau_1\} \infere e : \tau_2
}{
\Gamma \infere (\lambda x . e) : \tau_1 \fleche \tau_2}~(fun)
\quad\quad
\regle{
\Gamma \infere e_1 : \tau' \fleche \tau \quad \Gamma \infere e_2 : \tau'
}{
\Gamma \infere e_1~e_2 : \tau}~(app)$$

$$\regle{
\Gamma \infere e_1 : \tau_1 \quad
\Gamma\union\{x:Gen(\tau_1,\Gamma)\} \infere e_2:\tau_2
}{
\Gamma \infere (let ~ x = e_1 ~in~  e_2) : \tau_2}~(let)$$

$$\regle{
 \Gamma_i  \infere e_i : \col{\seq}{\tau'}  
\quad
 \Gamma_i  \infere g_i : \bool 
\quad
(1 \leq i \leq n)
}{
\Gamma \infere \{
m_1/g_1 \implique e_1 ;\dots; m_n/g_n \implique e_n \} : \col{\rho}{\tau} \fleche \col{\rho}{\tau'}
}(trans)$$

\medskip

où $\Gamma_i = \Gamma \union \{\self:\col\rho\tau\}\union \gamma(m_i,\tau)$
et  avec~$\gamma$ définie par $\gamma((x_1,\dots,x_k),\tau)=\{x_1:\tau,\dots,x_k:\tau\}$.\\

Comme usuellement la fonction~$Gen$ généralise un type~$\tau$ en
schéma de type~$\forall
\alpha_1,\dots,\alpha_n,\theta_1,\dots,\theta_m.\tau$ où les variables
de type et de topologie quantifiées sont les variables du type qui ne
sont pas liées dans le contexte de typage.

Dans la règle \emph{(trans)} on type toutes les règles comme si elles
avaient la même forme bien que la dernière règle n'ait pas de
garde. Ceci est naturel car le motif $x$ est équivalent au motif
$x/true$.
Intuitivement la règle \emph{(trans)} exprime le fait suivant : si la
partie droite de chaque règle est une séquence de type contenu $\tau'$
lorsqu'on suppose que les variables liées dans son motif ont le
type~$\tau$ alors la transformation renvoie une collection de type
contenu~$\tau'$ lorsqu'elle est appliquée à une collection de
type~$\tau$. Ceci est vrai grâce à la règle attrape-tout rendue
obligatoire par la syntaxe des transformations. La collection renvoyée
a la même topologie que la collection en argument.

\`A l'intérieur d'une transformation l'identificateur~$\self$ est
lié à la collection à laquelle la transformation est appliquée.  On
peut remarquer que si~$\self$ n'est pas utilisé dans le corps de la
transformation celle-ci pourra toujours être polytypique.

Si $\emptyset \infere e : \tau$ (noté aussi $\infere e : \tau$) alors
l'évaluation de $e$ ne provoquera pas d'erreur de type \wrong. En
revanche, l'absence d'erreur de structure newtonienne n'est pas
assurée.

\subsubsection*{Exemple}

On peut montrer avec les règles de typage que la transformation suivante a le type $\col\theta{\int}\fleche \col{\theta}{\int}$ pour toute topologie $\theta$~:
$\{ x, y/x>y \implique [x,  y, (x-y)] ; x \implique [x] \}$

La preuve est donnée ci-dessous avec 
$\Gamma_1=\{x:\int ;~ y:\int ;~ \self:\col\theta{\int}\}$ et
$\Gamma_2=\{x:\int ;~ \self:\col\theta{\int}\}$.

\newcommand{\racine}{\infere 
	\{ x,y/x>y \implique [x,y,(x-y)] ; x\implique[x] \} :
	 \col\theta{\int} \fleche \col{\theta}{\int}}

\newcommand{\rl}{\Gamma_1 \infere x>y : \bool}

\newcommand{\rr}{\Gamma_1 \infere [x,y,(x-y)] : \col{\seq}{\int}}

\newcommand{\rrl}{\Gamma_1 \infere  x :\int}
\newcommand{\rrr}{\Gamma_1 \infere  [y,(x-y)]  : \col{\seq}{\int}}

\newcommand{\rR}{\Gamma_2 \infere [x] : \col{\seq}{\int}}
\newcommand{\rRc}{\Gamma_2 \infere  x : \int}
\newcommand{\rRcc}{\Gamma_2( x) \leq \int}

%\begin{figure}[htbp]
%\medskip
\begin{center}

%\subfigure[]{
$
\regle{
\regle{\dots}{\rl} 
\quad
\regle{\regle{\Gamma_0( x) \leq \int}{\rrl} \quad \regle{\dots}{\rrr}}{\rr}
\quad
\regle{\regle{\rRcc}{\rRc}}{\rR}
}
{
\racine
}
$
%}
%
%

\end{center}

%\caption{Exemple de vérification de type}
%\label{tutu}
%\vspace{0.4cm}
%\hrule

%\end{figure}

\subsection{Inférence automatique}

L'algorithme d'inférence de type automatique $\mcal W$ de Damas/Milner
s'étend simplement au langage $\mcal L_=$ et aux règles de typage
correspondantes. Il suffit pour cela d'étendre la procédure
d'unification de Robinson aux types collections et aux topologies
ainsi que d'ajouter un cas dans~$\mcal W$ pour le typage des
transformations. Cet algorithme est donné dans~\cite{rule03entcs} et a
été implémenté afin d'être intégré à un compilateur pour une version
fortement typée du langage~MGS.  Comme~$\mcal W$, il calcule le type le
plus général d'un programme.

%%%%%%%%%%%%%%%%%%%%%%%%%%%%%%%%%%%%%%%%%%%%%%%%%%%%%%%%%%%%%%%
% TYPAGE SOUPLE %%%%%%%%%%%%%%%%%%%%%%%%%%%%%%%%%%%%%%%%%%%%%%%
%%%%%%%%%%%%%%%%%%%%%%%%%%%%%%%%%%%%%%%%%%%%%%%%%%%%%%%%%%%%%%%
\section{Typage souple}
\label{souple}

Le langage~$\mathcal L_=$ ne permet la manipulation des collections
que dans un cadre homogène. Nous présentons à présent le
langage~$\mathcal L_\subseteq$ qui est presque identique à~$\mathcal
L_=$ mais qui permettra des tests dynamiques de type afin de manipuler
des collections hétérogènes. Nous associons à ce langage un système de
types approprié qui effectue un typage statique tout en laissant
certains tests de types à l'exécution. Ce système de types plus avancé
que le premier utilise des types union et du sous-typage
non-structurel. La procédure d'inférence automatique est plus complexe
que celle basée sur~$\mathcal W$.

\subsection{Langage $\mcal L_\subseteq$}
\label{langage-souple}

Au niveau syntaxique $\mcal L_\subseteq$ diffère de~$\mcal L_=$ par la
possibilité de tester dynamiquement le type d'une valeur durant le
filtrage et la liberté d'avoir une règle attrape-tout dans la
transformation ou non. Les tests de type sont spécifiés par
annotation des variables des motifs comme le montre la syntaxe
ci-dessous. Ces conditions de types sont restreintes aux types de
base. La construction $\mu$ est appelée \emph{motif élémentaire} et
$b$ désigne un type de base de $B=\{\int,~\bool,~\float,~string\}$.
\begin{center}
\begin{tabular}{rcl}
$e$&$::=$&$x ~|~ \lambda x . e ~|~ e ~ e ~|~ c 
	     ~|~ \mbox{let }x=e\mbox{ in } e ~|~\{m/e \implique e;\dots;m/e\implique e\} $\\
 & & \\
$m$&$::=$ & $\mu,\dots,\mu  $\hfill$\mu$ ~ $::=$ ~ $x ~|~ x:b$

\end{tabular}\end{center}

\rem{Comme pour le langage~$\mcal L_=$, le langage~$\mcal L_\subseteq$ est
évalué dans le domaine~$D$, qui contient les valeurs particulières
$wrong$ et $shape\_err$. La valeur~$\bottom$ dans~$D$ dénote un calcul
qui ne termine pas. 
La définition de la fonction $eval$ sur les transformations est donnée
dans~\cite{tech89}. }

Le langage $\souple$ s'évalue dans un domaine $D$ qui contient les
collections topologiques en plus des valeurs usuelles
(voir~\cite{handbook} pour une introduction aux domaines sémantiques
et~\cite{tech61} ou~\cite{topological2002} pour le modèle des
collections topologiques). Les transformations sont représentées dans
$D$ par des fonctions continues\footnote{On peut considérer que les
transformations sont déterministes en supposant que la stratégie
d'application d'une règle est fixée mais non spécifiée.}. Leur
sémantique est donnée dans~\cite{tech89}.
$D$ contient les valeurs spéciales $\wrong$ et $\shaperr$ ainsi que $\bottom$ qui dénote un calcul qui ne termine pas.

\`A tout type de base $b$ on associe un sous-ensemble de
$D-\{wrong,shape\_err\}$ noté $\llbracket b \rrbracket$ et
contenant~$\bottom$. L'intersection de ces sous-ensembles deux à deux
doit toujours valoir $\{ \bottom \}$.

Un \emph{environnement} est une fonction d'un ensemble d'identifiants vers l'ensemble des valeurs $D$. On note $\Eval(e,E)$ la sémantique de l'expression $e$ dans l'environnement $E$.

% TYPES %%%%%%%%%%%%%%%%%%%%%%%%%%%%%%%%%%%%%%%%%%%%%%%%%%%%%%%%%%%%%%%%%%%%%

\subsection{Types}
\label{types}

La syntaxe des types est la suivante~:

\begin{tabular}{rcl}

$ \tau$ & ::= &$B ~|~\alpha ~|~\tau_1 \fleche \tau_2
   ~|~\col{\rho}{\tau} ~|~ \tau_1 \union \tau_2 ~|~ \tau_1 \inter
   \tau_2 ~|~ 0 ~|~ 1~|~ \tau_1?\tau_2$\\ & &\\ $\rho$ & ::= & $R ~|~
   \theta$
\end{tabular}

L'ensemble $R$ est le même que pour~$\mcal L_=$.
La sémantique des types est basée sur la notion
d'\emph{idéal}~\cite{ideaux}~: un type correspond à un sous ensemble
particulier du domaine des valeurs~$D$. La relation de sous-typage est
notée~$\subseteq$ et correspond à l'inclusion ensembliste sur~$D$. Un
type ne peut contenir ni~$wrong$ ni~$shape\_err$.

Voici une interprétation intuitive des types avant que nous ne
donnions leur sémantique formelle.
\begin{itemize}

\item
Le type \emph{flèche}, les types de base et les variables de type sont
interprétés comme usuellement dans les langages fonctionnels.

\item
Dans le type collection $\col\rho\tau$, le type contenu est~$\tau$ et la
topologie est~$\rho$ comme pour~$\mcal L_=$.

\item
Les types $\tau_1 \union \tau_2$ et $\tau_1 \inter \tau_2$
correspondent à l'union et l'intersection ensembliste des types.

\item
Le type 0 contient uniquement la valeur~$\bottom$, qui représente la
non-terminaison. Le type 0 est inclus dans tous les autres types
car~$\bottom$ appartient à tous les types. Le type 1 contient toutes
les valeurs sauf \emph{wrong} et~\emph{shape\_err}. Le type 1 inclut
tous les autres types. Le type $0\fleche 1$ convient à toute fonction,
y-compris aux transformations.

\item 
Le type conditionnel $\tau_1 ? \tau_2$ vaut $\tau_1$ lorsque $\tau_2$
est différent de $0$ et il vaut $0$ sinon. Par exemple le
type $\int?(\tau \inter \float)$ vaut $\int$ si $\tau$ contient
$\float$ et $0$ sinon.

\end{itemize}

Un schéma de type $\sigma$ est de la forme $\forall \alpha_1, \dots,
\alpha_n, \theta_1, \dots, \theta_m . \tau~where~S$ où~$S$ est un
ensemble de contraintes de types de la forme $\tau_1 \subseteq
\tau_2$. On utilisera par la suite le symbole~$\chi$ pour désigner
indifféremment une variable de type ou une variable de topologie. On
note $\Sol(S)$ l'ensemble des solutions de $S$.

\subsubsection*{Interprétation sémantique des types}

\'Etant donné une instanciation $s$ des variables de type et de
topologie, la sémantique d'un type et d'un schéma de types sont
définis dans la figure~\ref{semantiquetypes} par la fonction $\lcroch
. \rcroch_s$. 

La sémantique d'un type de base $b$ est l'ensemble $\lcroch b \rcroch$
défini en section~\ref{langage-souple}. La sémantique du type $\tau_1
\fleche \tau_2$ est l'ensemble des fonctions continues de~$D$ vers~$D$
telles que $f(v)$ est dans $\lcroch \tau_2 \rcroch_s$ (ou provoque une
erreur différente de \wrong) si $v$ est dans $\lcroch \tau_1
\rcroch_s$. La sémantique de $\tau_1 \union \tau_2$ est l'union de la
sémantique de $\tau_1$ et de la sémantique de $\tau_2$. La sémantique
du type $\col{\rho}{\tau}$ est l'ensemble des collections de~$D$ dont
la topologie correspond à~$\rho$ et dont les éléments sont dans
$\lcroch \tau \rcroch_s$.
La sémantique du schéma de types $\forall
\chi_1,\dots,\chi_n.\tau~where~S$ est l'ensemble des valeurs qui sont
dans $\lcroch \tau \rcroch_{s'}$ pour toute instanciation $s'$ des
$\chi_i$ solution de $S$ ($s'$ doit être compatible avec $s$ sur les
variables non quantifiées).

\begin{figure*}[tbp]
%\hrule 
\vspace{0.3cm}
\begin{center}
\begin{tabular}{rcl}
$\lcroch \alpha \rcroch _s$ &=& $\lcroch s(\alpha) \rcroch_\emptyset$\\
$\lcroch b \rcroch_s$ &=& $\lcroch b \rcroch$\\
$\lcroch \tau_1 \fleche \tau_2 \rcroch_s $ &=& 
$\big\{ f \in D \fleche D~|~ f(\lcroch \tau_1 \rcroch _s - \{\bottom\})
\subseteq \lcroch \tau_2 \rcroch_s \union \{shape\_err\} \big\} \union \{ \bottom \}$\\
$\lcroch \tau_1 \union \tau_2 \rcroch_s$ &=&
$\lcroch \tau_1 \rcroch_s \union \lcroch \tau_2 \rcroch_s$\\
$\lcroch \tau_1 \inter \tau_2 \rcroch_s$ &=&
$\lcroch \tau_1 \rcroch_s \inter \lcroch \tau_2 \rcroch_s$\\
$\lcroch \tau_1 ? \tau_2 \rcroch_s$ &=& $\bigg\lbrace$
\begin{tabular}{cl} 
$\lcroch\tau_1\rcroch_s$ &  si $\lcroch \tau_2 \rcroch_s \not = \{\bottom\}$\\
$\{\bottom\}$ & sinon
\end{tabular}\\
$\lcroch0\rcroch_s$ & = & $ \{\bottom\}$\\
$\lcroch1\rcroch_s$ & = & $D - \{wrong,shape\_err\}$\\
$\lcroch \col\rho\tau\rcroch_s$ & = & $\{c\in D ~|~ s(\rho)\mbox{ est la topologie de }c\mbox{ et }\forall e \in c.e\in\lcroch \tau \rcroch_s\}$  \\
 & & \\
$\lcroch \forall \alpha_1,\dots,\alpha_n,\theta_1,\dots,\theta_m.\tau~where~S\rcroch_s$ &=&
$\bigcap\limits_{s' \in X} \lcroch \tau \rcroch_{s'}$\\
 & & où $X = Sol(S) \inter$\\
 & & \phantom{où $X =$}$
\{ s' | s'(\chi) = s(\chi)
\mbox{ si }\chi \not\in \{\alpha_1,\dots,\alpha_n,\theta_1,\dots,\theta_m\}\}$
 
\end{tabular}

\caption{Sémantique des types et des schémas de types}
\label{semantiquetypes}
\end{center}
\hrule
\end{figure*}

\`A chaque constante $c$ du langage on associe un schéma de
types~$TC(c)$. On suppose que $TC$ est correct par rapport à la
sémantique~: pour toute constante $c$ du langage et pour toute
instanciation~$s$ des variables de type et de topologie,
$\Eval(c,\emptyset) \in \llbracket TC(c)\rrbracket_s$.
% RÈGLES D'INFÉRENCE %%%%%%%%%%%%%%%%%%%%%%%%%%%%%%%%%%%%%%%%%%%%%%%%%%%%%%%

\subsection{Règles de typage}

La figure~\ref{regles} donne notre système de règles de typage pour 
$\mcal L_\subseteq$. La relation de typage comporte un contexte $\Gamma$ et un
ensemble $S$ de contraintes de types. Le jugement $\Gamma,S \infere e :
\tau$ peut se lire {\og dans le contexte de typage~$\Gamma$,
l'expression~$e$ a le type $s(\tau)$ pour toute solution $s$ de
$S$ \fg}. La présence d'un ensemble de contraintes dans la relation de
typage est standard dans les systèmes de types en présence
de sous-typage. Toutes les règles sauf~\emph{(const)} et~\emph{(trans)} sont similaires à celles de Aiken~\emph{et al.}

Voici une description des
règles de la figure~\ref{regles}~:\begin{description}

\item[(var) :]
Cette règle correspond à la règle standard de Hindley/Milner.

\item[(const) :] 
$TC$ donne les schémas de types des constantes. 

\item[(fun) et (app) :] 
La règle \emph{(fun)} correspond à celle de
Hindley/Milner. Dans~\emph{(app)} les contraintes expriment qu'une
fonction du type $\tau_3 \fleche \tau_4$ ne peut être appliquée à une
valeur du type~$\tau_2$ que si~$\tau_2$ est un sous-type de~$\tau_3$.

\item[(gen) et (inst) :] La règle \emph{(gen)} sert à introduire le
polymorphisme paramétrique dans les types. En effet elle exprime que
si une expression a le type $\tau$ sous les contraintes $S$ alors elle
a aussi le schéma de type $\forall(\chi_i).\tau\mbox{ $where$ }S$ où
les $\chi_i$ sont des variables libres de $\tau$. La règle
\emph{(inst)} sert à instancier les schémas de types en types. Pour
utiliser cette règle, les contraintes du schéma de types doivent avoir
une solution. Les $\tau_i$ et $\rho_j$ sont libres dans cette règle.

\item[(let) :]
Le \emph{let-polymorphisme} est obtenu en
appliquant la règle~\emph{(gen)} juste après la règle~\emph{(let)}.

\item[(trans) :] Dans cette règle nous utilisons deux fonctions
définies inductivement sur les motifs : $\Comp$ et $\gamma$. La
première, $\Comp$ s'applique à un motif $m$ et à un type $\tau$ et
renvoie un type qui vaudra toujours~$\{\bottom\}$ si le motif ne peut
s'appliquer dans une collection de type contenu~$\tau$ et qui vaudra
un type différent de~$\{\bottom\}$ sinon. On dira que cette fonction
calcule la \emph{compatibilité} d'un motif avec un type. Par exemple
$\Comp~ ( (x_1\!:\!\int,~x_2\!:\!\float),\;\tau) = (\tau \inter
\int)\;?\; (\tau \inter \float)$. Ainsi si $\tau$ ne contient pas
$\int$ et $\float$ ce type vaut~$\{\bottom\}$.  La seconde fonction,
$\gamma$, calcule le contexte induit par un motif~$m$ sachant que ce
motif est appliqué à une collection de type contenu~$\tau$.

Pour une règle $m_i/g_i \implique e_i$ dans \emph{(trans)}, $\tau_i$
est le type contenu de la séquence~$e_i$ sachant que la transformation
s'applique à une collection de type contenu~$\tau$ et en considérant
le contexte induit par~$m_i$. Le type $\tau_i ? \Comp(m_i,\tau)$
vaudra~$\{ \bottom \}$ si les conditions de types de $m_i$ font que le
motif n'a jamais d'instances dans une collection de type
contenu~$\tau$ (incompatibilité), il vaudra $\tau_i$ si la règle peut
s'appliquer (compatibilité).

Le type contenu de la collection renvoyée doit être un sur-type de
$\tau_i ? \Comp(m_i,\tau)$ pour chaque~$i$, d'où les contraintes
$\tau_i ? \Comp(m_i,\tau) \subseteq \tau'$ car si la règle peut s'appliquer, la collection renvoyée pourra contenir des éléments de type $\tau_i$.
Par ailleurs lors de l'application d'une transformation des valeurs
peuvent ne pas être filtrées et resteront dans la collection renvoyée,
d'où la contrainte\footnote{\label{catchall}On peut ne pas considérer
cette contrainte lorsque l'on sait détecter que toutes les valeurs
seront filtrées, afin d'avoir un type plus précis. Par exemple
dans~$\mcal L_=$ la règle attrape-tout assure que tous les éléments
sont filtrés.}~$\tau \subseteq \tau'$.

\end{description}

\newcommand{\reglevar}{ {\over\Gamma \union \{ x : \sigma \},S \infere x : \sigma} ~(var)}

\newcommand{\reglefun}{{\Gamma \union \{x:\tau_1\},S \infere e: \tau_2 \over
\Gamma,S\infere \lambda x . e : \tau_1 \fleche \tau_2}~(fun)}

\newcommand{\regleconst}{{\over
 \Gamma,S \infere c : TC(c)}~(const)}

\newcommand{\regleapp}{{\Gamma,S \infere e_1 : \tau_1,~ e_2 : \tau_2 \over
\Gamma,S \union\{ \tau_2 \subseteq \tau_3,
 \tau_1 \subseteq \tau_3 \fleche \tau_4\}
\infere e_1~e_2 : \tau_4}~(app)}

\newcommand{\reglegen}{{\Gamma,S \infere e: \tau 
\over
\Gamma,\emptyset \infere e : \forall \chi_1,\dots,\chi_n.\tau~where~S} ~(gen) ~\mbox{si } Sol (S) \not = \emptyset \mbox{ et }
 \chi_1,\dots,\chi_n \mbox{ non libres dans } \Gamma}

\newcommand{\regleinst}{{\Gamma,S \infere e: \forall \alpha_1,\dots,\alpha_n,\theta_1,\dots,\theta_m . \tau~where~S' 
\over 
\Gamma,S \union S'[\tau_i/\alpha_i,\rho_j/\theta_j] \infere e : \tau[\tau_i/\alpha_i,\rho_j/\theta_j]}~(inst)}

\newcommand{\reglelet}{{ \Gamma,S \infere e_1:\sigma \quad \Gamma \union \{ x : \sigma \}, S \infere e_2 : \tau 
\over
\Gamma,S \infere \mbox{let } x = e_1 \mbox{ in } e_2 : \tau} ~(let)}

\newcommand{\regletrans}{\regle{
\Gamma_i, S \infere
g_i : \bool
\quad\quad 
\Gamma_i , S \infere
e_i : \col{\seq}{\tau_i}
\quad\quad (1 \leq i\leq n)
}{\Gamma,S \union S' \infere
\{m_1/g_1 \implique e_1 ; \dots ;m_n /g_n\implique e_n \} :
 \col\rho\tau \fleche \col{\rho}{\tau'}
}~(trans)}

\begin{figure*}[htb]
\hrule
\vspace{0.4cm}
$$ \reglevar \quad\quad\quad\quad \reglefun  $$

\medskip

$$\regleconst \quad\quad\quad\quad \regleapp$$ 

\medskip

$$\reglegen$$

$$\regleinst \quad\quad\quad \reglelet$$

\medskip

$$\regletrans$$

\medskip

où $S'=\{ \tau \stype \tau'\}\union \bigcup\limits_{1\leq i\leq
n} \{\tau_i ? \Comp(m_i,\tau) \stype \tau' \}$ ~~~et~~~ $\Gamma_i =
\Gamma \union\{\self : \col\rho\tau\} \union \gamma(m_i,\tau)$.\\

\bigskip

\noindent\begin{tabular}{ll}
$\Comp~(x,\;\tau) $&$= \tau \inter 1 = \tau $\\
$\Comp~(x:b,\;\tau) $&$= \tau\inter b$\\
$\Comp~((x,m'),\;\tau)  $&$=\Comp~(m',\tau)$\\
$\Comp~((x:b,m'),\;\tau)  $&$=(\tau\inter b)\;?\;\Comp\;(m',\tau)$
\end{tabular}\hfill
\begin{tabular}{ll}
$\gamma( x,~\tau)$ & $= \{x:\tau\}$\\
$\gamma( x:b,~\tau)$ & $= \{x:\tau\inter b\}$\\
$\gamma( (x,m'),~\tau)$ & $= \{x:\tau\}\union \gamma (m',~\tau)$\\
$\gamma( (x:b,m'),~\tau)$ & $= \{x:\tau\inter b\}\union \gamma (m',~\tau)$\\
\end{tabular}

\caption{Règles de typage}
\label{regles}

\vspace{0.3cm}
\hrule

\end{figure*}

\subsubsection*{Exemples}

\newcommand{\lignei}
{\{\self : \col\theta\tau ; x : \alpha \inter \int \}, \emptyset 
	\infere [x;1] : \col{\seq}{\int}}

\newcommand{\ligneii}
{ \emptyset, \{\alpha \stype \alpha ,  (\int ? (\alpha \inter \int))\subseteq \alpha\}
 \infere \{x:\int \implique [x;1] \} : \col\theta\alpha \fleche \col\theta\alpha}

\newcommand{\ligneiii}
{\emptyset,\emptyset \infere \{x:\int \implique [x;1] \} :
\forall \alpha, \theta.\col\theta\alpha \fleche \col\theta\alpha~where~
\{\alpha \stype \alpha ,  (\int ? (\alpha \inter \int))\subseteq \alpha\}}

Voici les preuves de typage de deux transformations simples.
$$ \regle{\regle{\regle{\dots}{\lignei} }{\ligneii}}{\ligneiii}$$ 

Dans cet exemple l'ensemble de contraintes du schéma se réduit à
$\emptyset$. En effet on peut montrer que $(\int ? (\alpha \inter
\int))\subseteq
\alpha$ est toujours vrai. Donc n'importe quelle instanciation
de~$\alpha$ et~$\theta$ convient.

\newcommand{\lignej}
{\{\self : \col\theta\tau ; x : \alpha \inter \int \}, \emptyset 
	\infere [true] : \col{\seq}{\bool}}

\newcommand{\lignejj}
{ \emptyset, \{\alpha \stype \beta ,  (\bool ? (\alpha \inter \int))\subseteq \beta\}
 \infere \{x:\int \implique [true] \} : \col\theta\alpha \fleche \col\theta\beta}

\newcommand{\lignejjj}
{\emptyset,\emptyset \infere \{x:\int \implique [true] \} :
\forall \alpha,\beta, \theta.\col\theta\alpha \fleche \col\theta\beta~where~
\{\alpha \stype \beta ,  (\bool ? (\alpha \inter \int))\subseteq \beta\}}

$$ \regle{\regle{\regle{\dots}{\lignej} }{\lignejj}}{\lignejjj}$$ 

Ici le type $\alpha \union (\int ? (\alpha \inter \int))$ est la
plus petite instanciation de $\beta$ vérifiant les contraintes du schéma. Le
type suivant est donc valable pour cette transformation~:
\mbox{$\col\theta\alpha \fleche \col\theta{\alpha \union (\bool ? (\alpha
\inter \int))}$}.

Sans l'utilisation de types conditionnels, le type le plus précis pour
cette transformation aurait été $\col\theta\alpha \fleche
\col\theta{\alpha \union \bool }$ qui porte moins d'informations que le
type précédent.\\

\subsection{Propriétés}

%Un environnement est une fonction d'un ensemble de variables du
%langage vers l'ensemble des valeurs. 
On dira qu'un environnement $E$ est correct par rapport à un
contexte~$\Gamma$ et une instanciation~$s$ des variables de type et de
topologie lorsque $E(x) \in \llbracket \Gamma(x)\rrbracket_s$ pour
tout $x$ lié dans $\Gamma$ et $E$.

\begin{lemme}[Correction]
\label{correction}
Soit un typage $\Gamma,S\infere e : \sigma$, une solution~$s$ de $S$,
un environnement $E$ correct par rapport à~$\Gamma$ et~$s$ portant sur les variables libres de $e$.
Alors~$ \Eval(e,E)\in \lcroch \sigma \rcroch_s \union \{shape\_err\}$.

\end{lemme}

La preuve est donnée dans~\cite{tech89}.  Le corollaire suivant
découle de ce lemme~: si~$e$ est une expression sans variables libres
et si $\emptyset,\emptyset \infere e : \sigma$ alors
$Eval(e,\emptyset) \not = wrong$. Ceci est vrai car $wrong$
n'appartient à aucun type.  On dit qu'un programme~$e$ est \emph{bien
typé} si il existe un schéma de type~$\sigma$ tel que
$\emptyset,\emptyset\infere e : \sigma$. Notons que le lemme ne
garantit rien sur les erreurs de structure newtonienne.\\

% INFÉRENCE AUTOMATIQUE %%%%%%%%%%%%%%%%%%%%%%%%%%%%%%%%%%%%%%%%%%%%%%%%%%%%%

\subsection{Algorithme d'inférence automatique}
\label{auto}

L'inférence automatique des types d'un programme consiste en deux
étapes. En premier lieu, le schéma de type le plus général du
programme est calculé en appliquant les règles de typage selon une
stratégie appropriée. Ensuite on calcule les solutions des contraintes
de types générées. Nous décrivons ces deux étapes dans cette section.

\subsubsection{Production du type et des contraintes}

Nous suivons la stratégie d'application des règles proposée
  par~\cite{aiken93} qui définit la dérivation la plus générale modulo
  renommage des variables de type et de topologie~:
\begin{itemize}

\item 
On utilise des variables fraîches partout où cela est possible.

\item
On applique la règle~\emph{(gen)} immédiatement après la règle~\emph{(let)}.

\item
On applique~\emph{(inst)} après avoir appliqué la règle~\emph{(var)}
ou la règle~\emph{(const)}.

\item
La dérivation se termine par une application de la règle~\emph{(gen)}.

\item
Les règles \emph{(gen)} et \emph{(inst)} ne sont appliquées nulle
part ailleurs. 

\end{itemize}

\subsubsection{Résolution des contraintes}

Une procédure de résolution de systèmes de contraintes ensemblistes
est donnée dans~\cite{aiken93} et~\cite{aw94}. Elle est basée sur un
système de réécriture qui met les contraintes sous une forme où leurs
solutions peuvent être directement extraites. Cette procédure s'étend
aux types collections en considérant l'équivalence suivante que l'on
orientera de gauche à droite et en résolvant les égalités entre
topologies par unification~:
$\{ \col{\rho_1}{\tau_1} \subseteq \col{\rho_2}{\tau_2} \} \equiv \{
\rho_1 = \rho_2 ~;~ \tau_1 \subseteq \tau_2 \}$.
 
La présence d'opérateurs typés dans notre langage impose également de
modifier l'une des règles de réécriture de la procédure d'Aiken
\emph{et al.} : $\{\tau_1 \fleche \tau'_1 \subseteq \tau_2 \fleche
\tau'_2\}$ se réécrit en $\{ \tau_1 \subseteq \tau'2~;~\tau_2
\subseteq \tau'1\}$. 

La procédure de résolution proposée n'est correcte que sous une
condition sur la forme du système à résoudre, laquelle est détaillée
dans~\cite{tech89}. Nous montrons dans ce même document que notre
procédure de production de contraintes ne produit que des systèmes
solvables par notre procédure de résolution. On peut noter que cette
condition empêche de donner d'utiliser des types intersections pour
dénoter la surcharge d'opérateurs. Cependant nous verrons en
section~\ref{extensions} que la surcharge peut s'exprimer autrement
dans notre système de types.

\rem{
Un système est solvable par cette procédure et on peut en construire
toutes les solutions si et seulement si ses contraintes sont dans une
forme dite~\emph{convenable}.

Les contraintes produites par notre algorithme sont convenables
lorsqu'on contraint $TC$ à ne donner que des types convenables à leur
tour. Cette restriction sur $TC$ est décrite dans~\cite{tech89} et
consiste essentiellement à limiter l'usage des types union et
intersection. Ceci n'empêche pas de typer les opérateurs usuels des
langages fonctionnels, on peut avoir par exemple le type $\bool\fleche
\alpha \fleche \beta \fleche (\alpha\union \beta)$ pour
l'opérateur~$if$, en revanche on ne peut coder la surcharge avec une
intersection~: on ne peut avoir le type $(\int\fleche \int \fleche
\int)\inter (\float \fleche \float \fleche \float)$ pour l'opérateur~$+$
car les types intersections entre des types~\emph{flèche} ne peuvent
apparaître dans la partie gauche d'une contrainte.}

% COMPILATION %%%%%%%%%%%%%%%%%%%%%%%%%%%%%%%%%%%%%%%%%%%%%%%%%%%%%%%%%%%%%%%

\subsection{Compilation}
\label{compilation}

En plus du gain en performances attendu lorsqu'on passe d'un langage
dynamiquement typé à un langage statiquement typé, le typage de
$\mathcal L_\subseteq$ apporte une information pouvant s'apparenter à
de l'analyse de flot de contrôle, permettant des optimisations dans le
processus d'application des transformations. Dans cette section, nous esquissons certaines de ces optimisations. 

Considérons la règle $x:\int=>[x+1]$ et une collection~$c$ de type
contenu~$\tau$. Deux cas particuliers peuvent se présenter~:
\begin{itemize}

\item
Si d'après le type $\tau$ la collection $c$ ne contient pas d'entiers
alors la règle ne peut s'appliquer. De manière générale, une règle
$m/g\implique e$ ne peut s'appliquer si $\Comp(m,\tau)$ vaut
$\{\bottom\}$. On sait donc dès la compilation qu'il est inutile
d'essayer d'appliquer cette règle et on peut donc la sauter
(élimination des règles inutiles).

\item 
Si $\tau=\int$ alors on sait que \rem{toutes} les valeurs de la collection
vérifieront la condition de type du motif. Les tests de type sont donc
inutiles à l'exécution (élimination des conditions de types inutiles).
\end{itemize}

Supposons à présent que les collections soient implémentées de manière
à optimiser la recherche d'éléments lorsque leur type est connu. Par
exemple les ensembles peuvent être implémentés par des sous-ensembles
homogènes. Alors une optimisation est possible et elle généralise les
deux premières : on peut chercher les instances d'un motif élémentaire
dans la partie appropriée de la collection.

Les deux premières optimisations sont simples à implémenter mais des
études sont nécessaires pour savoir si elles s'appliquent souvent dans
les programmes réels. La dernière s'applique plus souvent mais il peut
être difficile d'implémenter les collections de manière à favoriser à la fois
la recherche en fonction du type et en fonction de la
topologie de la collection. En effet le processus de filtrage est
profondément lié à la topologie des collections puisqu'un motif filtre
des valeurs voisines.

% CONCLUSION %%%%%%%%%%%%%%%%%%%%%%%%%%%%%%%%%%%%%%%%%%%%%%%%%%%%%%%%%%%%%%%%
\section{Conclusion}
\label{conclusion}

\subsection{Comparaison des deux approches}

Comme le montre~\cite{wand}, l'algorithme~$\mcal W$ de Damas/Milner est
équivalent à un algorithme procédant par production de contraintes
d'égalités entre types suivie de résolution du système d'équations par
unification de Robinson.
Par conséquent les moteurs d'inférence automatique pour~$\fort$ et
pour~$\souple$ peuvent être implémentés en suivant un même schéma
production/résolution.
Pour passer de la production de contraintes pour~$\souple$ à la
production de contraintes pour~$\fort$ il s'agit essentiellement de
remplacer les inclusions par des égalités dans les
règles~\emph{(trans)} et~\emph{(app)}.
Par ailleurs l'équivalence $\{ X = Y \} \equiv \{ X \subseteq Y,~ Y \subseteq X \}$ montre que l'on peut mêler dans un
même système de contraintes des inclusions et des égalités. Une
stratégie d'application de l'unification et de la résolution
de Aiken~\emph{et al.} permet de résoudre de tels systèmes.
%
%$$
%
Ces deux observations nous on conduit naturellement à implémenter les
deux systèmes de types dans un seul moteur d'inférence automatique, en
cours d'intégration dans un compilateur MGS expérimental.

% EXTENSIONS %%%%%%%%%%%%%%%%%%%%%%%%%%%%%%%%%%%%%%%%%%%%%%%%%%%%%%%%%%%%%%%

\subsection{Extensions}
\label{extensions}

Nous présentons ici des extensions directes de nos travaux. Certaines
font déjà partie de notre implémentation du système alors que d'autres
sont des voies pour des travaux futurs.

\begin{description}

\item[Constructions usuelles.]
%\subparagraph{Constructions usuelles.}
%\textbf{Constructions usuelles.}
Les constructions usuelles telles que le produit ou le \emph{let-rec}
n'ont pas été considérées pour ne pas alourdir la présentation mais leur
ajout au langage et au typage se fait de manière habituelle et ils
sont présents dans notre implémentation.

\item[Gardes dans le motif.]
%\textbf{Gardes dans le motif.}
Pour la simplicité de la présentation nous avons restreint l'usage de
la garde dans un motif mais on peut étendre aisément le langage des
motifs afin d'avoir une garde associée à chaque motif élémentaire,
comme fait dans~\cite{rule03entcs} et garder les propriétés du typage fort
ou du typage souple. Mettre des gardes au plus tôt dans le motif
permet d'optimiser le processus de filtrage. Par exemple le filtrage
du motif $(x/x\!>\!0), y$ est plus rapide que pour $x,y/x\!>\!0$.
Cette extension fait partie de notre implémentation.
\rem{En effet, considérons une collection d'entiers de cardinal~$N$
dans laquelle aucune valeur n'est supérieure à~$0$, le processus de
filtrage du premier motif testera $N$~fois la garde avant d'échouer
tandis qu'il faudra $N*(N-1)$~tests pour le second.}

\item[Des types plus précis.]
%\textbf{Des types plus précis.}
Nous avons vu que le système de types de $\souple$ permettait
d'inférer des types assez précis. Pourtant dans au moins deux cas que
nous montrons ici des types plus précis existent.
\begin{itemize}

\item
Le type inféré pour la transformation $\{x:\int \implique [true]\}$
est $\col\theta\alpha \fleche \col\theta{\alpha \union (\bool ? (\alpha
\inter \int))}$. Ce type ne porte pas l'information que la collection
renvoyée ne contient plus d'entiers.

\item
Le type inféré pour la transformation $\{ x:\int \implique [x] ; x:\int
\implique [true] \} $ est $\col\theta\alpha \fleche
\col\theta{\alpha\union(\bool?(\alpha\inter \int))}$. Or la deuxième règle
ne s'applique jamais car tous les entiers sont consommés par la
première donc le type $\col\theta\alpha \fleche \col \theta \alpha$
qui est plus précis convient.

\end{itemize}
La règle~\emph{(trans)} de~$\souple$ analyse sommairement les
conditions dans lesquelles les règles peuvent s'appliquer (par le
biais des types conditionnels et de la fonction \Comp). Les deux
exemples ci-dessus montrent que l'analyse des motifs et des
transformations doit être plus subtile pour inférer le type le plus
précis d'un programme.  Par exemple l'amélioration proposée dans la
note de bas de page numéro~\ref{catchall} permet d'obtenir le typage
le plus précis pour la fonction~\emph{map} qui est $(\alpha \fleche
\beta) \fleche \col\theta\alpha \fleche \col\theta\beta$ au lieu de
$(\alpha \fleche \beta) \fleche \col\theta\alpha \fleche
\col\theta{\alpha\union\beta}$.

\item[\'Etoile dans un motif.] 
%\textbf{\'Etoile dans un motif.}
Les motifs tels que nous les avons présentés permettent uniquement de
filtrer des parties de taille prédéterminée. Il existe toutefois des
cas où le programmeur aimerait filtrer des parties de la collection de
taille arbitraire, comme montré dans~\cite{rule02}. Afin de permettre
ceci, nous introduisons une nouvelle construction dans les motifs
appelée l'\emph{étoile} et notée~$*$ qui exprime le filtrage d'un
nombre d'éléments arbitraire. La grammaire des motifs élémentaires est
modifiée comme suit~:
$$\mu ~::=~x ~|~ x:b ~|~ *~as~x ~|~ b\!*~as~x$$ 
Dans le motif élémentaire $*~ as~x$, l'identificateur~$x$ dénote la séquence des valeurs
filtrées par l'étoile et peut être utilisé dans la garde du motif et
dans l'expression de remplacement de la règle. Le fait que $x$ dénote
une séquence plutôt qu'une partie de la collection est en accord
avec la contrainte d'avoir une séquence comme expression de
remplacement.

Le typage de ces nouveaux motifs nécessite uniquement la modification
des fonctions~$\gamma$ et $\Comp$ de la manière suivante~:\begin{center}\begin{tabular}{ll}
$\gamma(*~as~x,\tau)$ & $=\{x:\col{\seq}{\tau}\}$\\
$\gamma(b\!*~as~x,\tau)$ & $=\{x:\col{\seq}{\tau\inter b}\}$
\end{tabular}
\hspace{1cm}
\begin{tabular}{ll}
$\Comp(*~as~x,\tau)$ & $=1$\\
$\Comp(b\!*~as~x,\tau)$ & $=\tau \inter b$
\end{tabular}\end{center}
Les répétitions arbitraires sont intégrées à notre implémentation.

\item[Direction à la place de la virgule dans un motif.] 
%\textbf{Direction à la place de la virgule dans un motif.}
Dans une structure de données comme la grille, on peut vouloir filtrer
des éléments voisins selon une direction donnée. Ceci peut ce faire
simplement comme dans le motif suivant~: \mbox{$x,y / (\nordnb~
\self~ x~ y)$} où~$\nordnb$ est la relation de voisinage
correspondant au constructeur~$\nord$. Toutefois il est
avantageux de placer cette information de voisinage au niveau
syntaxique afin de l'utiliser efficacement durant le filtrage. Ainsi
dans le langage MGS on écrit directement $x~|\nord\!>y$. La virgule a
été remplacée par le constructeur $\nord$ encadré des symboles~$|$
et~$>$. Cette nouvelle construction dans le langage ne nécessite pas
de modification dans le typage puisqu'elle peut être vue comme du
sucre syntaxique.

\rem{
\emph{Exemple.} Le tri par boulier est un méthode originale pour trier
des entiers écrits en une colonne de nombres en base
unaire~\cite{beadsort}. Les nombres sont triés en laissant tomber les
bits verticalement, comme le montre le schéma ci-dessous où l'on trie
les nombres [3,2,4,2]. Le premier schéma représente les nombres avant
le tri et le second après le tri. Les troisième et quatrième schémas
correspondent à l'implémentation des deux premiers sur une grille de
booléens où $true$ représente un bit et $false$ représente l'absence
de bit. La transformation suivante permet d'implémenter ce tri~:
\mbox{$\{x~|\nord\!>y / (x\!=\!false ~\&\&~ y\!=\!true) => [y,x]\}$}.
Le tri est obtenu en itérant l'application de cette transformation
jusqu'à l'obtention d'un point fixe. On peut montrer que cette
transformation a le type $\col{\grid}{\bool} \fleche\col{\grid}{\bool}$
avec $TC(\nordnb)=\col{\grid}\alpha\fleche \alpha \fleche \alpha
\fleche \bool$ et $TC(=) = \alpha \fleche \alpha \fleche \bool$. On peut
noter que cette transformation n'est pas polytypique car le type
de~$\nordnb$ contraint la collection argument à avoir la
topologie~$\grid$.

\begin{center}
\includegraphics[scale=0.5]{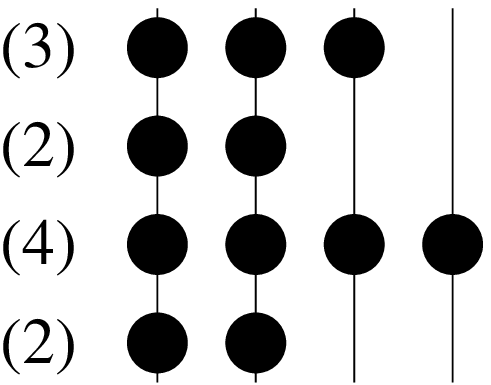}\hspace{0.7cm}
\includegraphics[scale=0.5]{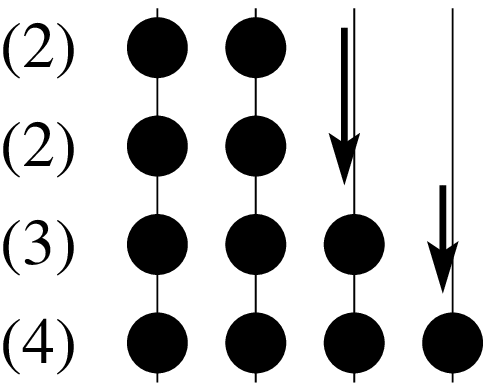}\hspace{0.7cm}
\includegraphics[scale=0.5]{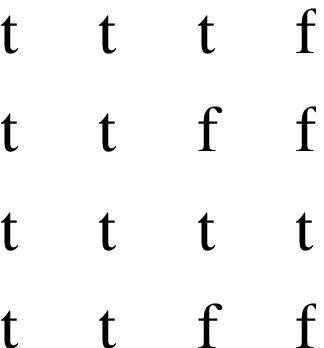}\hspace{0.7cm}
\includegraphics[scale=0.5]{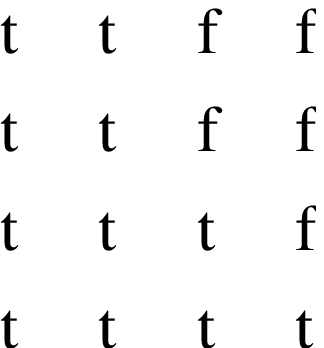}
\end{center}
}

\item[Lisibilité des types.]
%\textbf{Intelligibilité des types.}
Un schéma de types de $\souple$ contient un ensemble de contraintes
qui peuvent se révéler obscures pour le programmeur. F. Pottier a
formalisé une notion de simplification de systèmes de contraintes
dans~\cite{thesePottier} dans le but de rendre les schémas de types
plus lisibles par le programmeur. Une telle simplification nous semble
essentielle dans un but d'aide au développement.

\item[Surcharge.] 
%\textbf{Surcharge.}
%
Comme le font remarquer Pantel et Sallé dans~\cite{fol} les types
conditionnels introduits par Aiken~\emph{et al.} peuvent servir à
dénoter la surcharge de la manière suivante. Si~$\{U_i \fleche V_i\}$
est l'ensemble des types d'une fonction surchargée alors on peut lui
donner le type suivant dans notre système~:
$$\forall \alpha . \alpha \fleche \bigcup_i(V_i ? \alpha \inter
U_i)~where~\{\alpha \subseteq \bigcup_i U_i\}$$
\emph{Exemple.}  On peut donner le type suivant à l'addition sur les
entiers et les flottants~: $$\forall \alpha.\alpha \fleche
\alpha\fleche (\int ? \alpha \inter \int) \union (\float?\alpha \inter
\float)~where~\{\alpha \subseteq \int \union \float\}$$

\item[Types récursifs.]
%\textbf{Types récursifs.}
Les types récursifs ne sont pas intégrés directement à la grammaire
des types mais on peut définir des types récursifs à l'aide de
contraintes de la manière suivante~: le schéma de
types~$\forall\alpha.\col{\bag}{\alpha}~where~\{\alpha=\col{\bag}{\alpha}\union
\int\}$ dénote les multi-ensembles pouvant contenir des entiers et des
multi-ensembles contenant à leur tour des entiers et des
multi-ensembles et ainsi de suite. Dans cet exemple on a utilisé
l'égalité entre types définie par $\{\tau_1=\tau_2\} \equiv \{\tau_1
\subseteq \tau_2,\tau_2 \subseteq
\tau_1 \}$.

\end{description}

% REMARQUES %%%%%%%%%%%%%%%%%%%%%%%%%%%%%%%%%%%%%%%%%%%%
\subsection{Discussions}
\label{discussion}

\subparagraph{Erreurs de structure newtonienne.}
La faiblesse de nos systèmes de types est qu'ils ne permettent pas de
détecter les violations de structures newtoniennes à la
compilation. Pour le faire il faudrait garantir que le chemin filtré
et la séquence remplaçante ont la même taille. Or ceci n'est pas
toujours possible car~:\begin{itemize}

\item
la taille de la séquence remplaçante ne peut pas toujours être
calculée à la compilation~;

\item
la taille du motif n'est pas connue à la compilation si une répétition
arbitraire~$*$ y est utilisée.

\end{itemize}

Toutefois il existe des cas où l'on peut détecter que $shape\_err$
sera renvoyé ou ne sera pas renvoyé. Déterminer l'ensemble des
programmes pour lesquels l'apparition d'une $shape\_err$ est décidable
est souhaitable mais notre système de types est mal adapté à ceci. Il
existe également des systèmes de types prévus pour l'inférence automatique de
tailles~\cite{atable, size}.

\subparagraph{Types sommes ou types unions?}
Les types unions et les tests de type à l'exécution peuvent sembler
superflus lorsque l'on sait que les types sommes avec constructeurs
permettent de programmer de manière similaire dans un contexte
statiquement typé.  Les collections hétérogènes et les types unions
sont bien plus flexibles que les types sommes mais permettent moins de
vérifications automatiques, lesquelles sont essentielles à la
construction de logiciels sûrs. Par exemple ils ne permettent pas de
vérifier l'exhaustivité du filtrage, contrairement aux type sommes
(voir~\cite{maranget}).

Toutefois, dans certains domaines d'application, la complexité des
processus modélisés est telle que les types sommes deviennent trop
lourds. D'autres auteurs partagent cette idée. Par exemple J. Garrigue
étend les types sommes aux types sommes polymorphes dont les types ne
sont pas nécessairement déclarés et dont les constructeurs peuvent
être utilisés dans différents types~\cite{garrigue98}. Il les utilise
intensivement pour interfacer des bibliothèques C (comme OpenGL ou
GTK) au langage O'CAML. Les types sommes polymorphes apparaissent dans
la version 3 de O'CAML~\cite{ocaml}. Par ailleurs A. Frisch \emph{et
al.} utilisent des types unions pour typer le langage CDUCE dédié à la
manipulation de données XML~\cite{frisch02semantic}. Il existe
d'autres exemples où la manipulation de données venant de l'extérieur
nécessite un langage supportant l'hétérogénéité mais nous sommes
particulièrement motivés par des programmes issus de la biologie tels
que la simulation de cellules. Dans ce type de simulation on doit
gérer des entités qui migrent entre compartiments et des réactions
pouvant avoir lieu sans que l'ensemble des entités pouvant être
rencontrées ne soit connu \emph{a priori}. La définition d'un type somme pour
représenter l'ensemble des entités possibles est très lourde dans ce
cas. Les collections hétérogènes sont ici une solution appropriée. De
telles simulations ont été programmées à l'aide du langage MGS
(voir~\cite{bugrim}).

On peut également noter que l'approche orientée objet se prête à
l'implémentation de collections hétérogènes. Toutefois, que ce soit
avec des \emph{Vector} en JAVA ou des listes en O'CAML l'utilisateur
est obligé à un moment ou à un autre de coercer explicitement les
éléments des collections. Par ailleurs cette approche contraint à
considérer toujours des objets alors que bien souvent des types de
base suffisent. Enfin, la définition d'opérations polytypiques par le
programmeur reste assez lourde dans ce cadre.

\subsection{Travaux apparentés et perspectives}

Les travaux de Aiken \emph{et al.}~\cite{aiken93, aw94} fournissent
une méthode d'inférence de types en présence de types unions qui sert
de base à notre système de types souple, en effet les types union sont
la clé de notre représentation de l'hétérogénéité. Le filtrage
apparaissant dans~\cite{aw94} est en revanche prévu pour des termes et
n'est pas adapté à notre filtrage. Toutefois les types conditionnels
apparaissent déjà dans ces travaux pour exprimer une analyse de flots
lors du filtrage. Notre travail se démarque fortement du leur par le
point de vue original sur le typage des structures de données et la
puissance du processus de filtrage que nous typons.

Les travaux de A. Frisch \emph{et al.}~\cite{frisch02semantic} sont
assez proches des nôtres puisqu'ils proposent un système de types basé
sur un modèle ensembliste pour un langage avec filtrage sur le type
des valeurs où l'hétérogénéité des données est représentée par des
types unions. Toutefois la puissance de leurs motifs les contraint à
imposer une déclaration du type des fonctions
dans~\cite{frisch02semantic}.

D'autres systèmes de types existent pour des langages à base de
règles. Par exemple les travaux de P. Fradet \emph{et
al.}~\cite{structuredgamma} munissent le langage Gamma d'un système de
types dédié à la vérification de propriétés des programmes. En effet
Gamma est à l'origine un langage dédié à la spécification formelle de
haut niveau. Dans une optique d'implémentation efficace de $\souple$
notre système semble plus approprié. Le système de types de Gamma
permet en revanche d'envisager des vérifications de préservations de
topologie (sans résoudre les difficultés discutées en
section~\ref{discussion}).

Nous avons utilisé nos deux systèmes de types dans un compilateur MGS
 pour éliminer l'étiquetage des valeurs par leur type lorsque
 possible. Ceci permet de grandes améliorations des performances, ce
 qui nous incite à intégrer d'autres optimisations basées sur les
 types telles que l'élimination des règles inutiles (voir
 section~\ref{compilation}). D'autres langages à base de règles de
 réécriture pourraient tirer bénéfice de ces systèmes de types.

De nombreux travaux s'attellent à rendre le filtrage plus
expressif. Citons TOM~\cite{tom} un compilateur de filtrage adapté à
plusieurs langages (C, JAVA, Eiffel) et permettant le filtrage
associatif; ELAN~\cite{elan}, un langage fondé sur la réécriture et
proposant une implémentation très efficace du filtrage associatif et
commutatif; enfin CDUCE~\cite{frisch02semantic} et
G'CAML~\cite{furuse} qui permettent de filtrer les valeurs en fonction
de leur type. Les langages $\fort$ et $\souple$ qui sont des
sous-ensembles du langage MGS proposent du filtrage associatif,
commutatif, sur des structures non algébriques et également sur le
type des valeurs dans le cas de $\souple$.

Enfin, notre expérience de la programmation en MGS montre que les
enregistrements extensibles~\cite{remy} sont particulièrement utiles
et adaptés à une utilisation dans un cadre hétérogène (voir de
nombreux exemples de programmes dans~\cite{graphicgallery}). Nous
souhaitons ajouter les enregistrements à nos systèmes de types et
permettre les tests de types d'enregistrements lors du filtrage. Nous
pensons également étudier comment augmenter les tests de types dans le
filtrage sans perdre l'inférence automatique.

\subsection{Remerciements}
Je remercie Jacques Garrigue, Pascal Fradet, Catherine Dubois,
Giuseppe Castagna et Alain Frisch pour le temps qu'ils m'ont
consacré. Je remercie également Olivier Michel et Jean-Louis Giavitto
pour leur aide et l'équipe SP\'ECIF du LaMI pour leurs encouragements
constants.

Page web du projet MGS : \url{http://mgs.lami.univ-evry.fr}. Une page
est notamment dédiée aux travaux présentés dans cette article et aux
implémentations correspondantes:
\url{http://mgs.lami.univ-evry.fr/TypeSystem/}.

%////////////////////////////////////////////////////////////////////////biblio

%\bibliographystyle{alpha}

%\bibliography{BIB/biblio}

%///////////////////////////////////////////////////////////////////////////fin

\pagebreak
\thispagestyle{colloquetitle}
\cleardoublepage
\end{document}